\documentclass[iop]{emulateapj}
\let\pwiflocal=\iffalse \let\pwifjournal=\iffalse

\usepackage{amsmath,amssymb}
\usepackage[breaklinks,colorlinks,urlcolor=blue,citecolor=blue,linkcolor=blue]{hyperref}

\usepackage[T1]{fontenc}
\pwifjournal\else
  \usepackage{microtype}
\fi

\pwifjournal\else
  \makeatletter
  \renewcommand\plotone[1]{%
    \centering \leavevmode \setlength{\plot@width}{0.99\linewidth}
    \includegraphics[width={\eps@scaling\plot@width}]{#1}%
  }%
  \makeatother
\fi

\pwifjournal\else
  \makeatletter
  \long\def\@makecaption#1#2{%
    \par
    \vskip\abovecaptionskip
    \begingroup
     \small\rmfamily
     \sbox\@tempboxa{%
      \let\\\heading@cr
      \footnotesize #1 #2
     }%
     \@ifdim{\wd\@tempboxa >\hsize}{%
      \begingroup
       \samepage
       \flushing
       \let\footnote\@footnotemark@gobble
       \footnotesize #1 #2\par
      \endgroup
     }{%
       \global \@minipagefalse
       \hb@xt@\hsize{\hfil\unhbox\@tempboxa\hfil}%
     }%
    \endgroup
    \vskip\belowcaptionskip
  }%
  \makeatother
\fi

\pwifjournal\else
\fi

\makeatletter

\newcommand\@simpfx{http://simbad.u-strasbg.fr/simbad/sim-id?Ident=}

\newcommand\MakeObj[4][\@empty]{
  \pwifjournal%
    \expandafter\newcommand\csname pkgwobj@c@#2\endcsname[1]{\protect\object[#4]{##1}}%
  \else%
    \expandafter\newcommand\csname pkgwobj@c@#2\endcsname[1]{\href{\@simpfx #3}{##1}}%
  \fi%
  \expandafter\newcommand\csname pkgwobj@f#2\endcsname{#4}%
  \ifx\@empty#1%
    \expandafter\newcommand\csname pkgwobj@s#2\endcsname{#4}%
  \else%
    \expandafter\newcommand\csname pkgwobj@s#2\endcsname{#1}%
  \fi}%

\newcommand\MakeTrunc[2]{
  \expandafter\newcommand\csname pkgwobj@t#1\endcsname{#2}}%

\newcommand{\obj}[1]{%
  \expandafter\ifx\csname pkgwobj@c@#1\endcsname\relax%
    \textbf{[unknown object!]}%
  \else%
    \csname pkgwobj@c@#1\endcsname{\csname pkgwobj@s#1\endcsname}%
  \fi}
\newcommand{\objf}[1]{%
  \expandafter\ifx\csname pkgwobj@c@#1\endcsname\relax%
    \textbf{[unknown object!]}%
  \else%
    \csname pkgwobj@c@#1\endcsname{\csname pkgwobj@f#1\endcsname}%
  \fi}
\newcommand{\objt}[1]{%
  \expandafter\ifx\csname pkgwobj@c@#1\endcsname\relax%
    \textbf{[unknown object!]}%
  \else%
    \csname pkgwobj@c@#1\endcsname{\csname pkgwobj@t#1\endcsname}%
  \fi}

\makeatother

\pwifjournal\else
  \usepackage{etoolbox}
  \makeatletter
  \patchcmd{\NAT@citex}
    {\@citea\NAT@hyper@{%
       \NAT@nmfmt{\NAT@nm}%
       \hyper@natlinkbreak{\NAT@aysep\NAT@spacechar}{\@citeb\@extra@b@citeb}%
       \NAT@date}}
    {\@citea\NAT@nmfmt{\NAT@nm}%
     \NAT@aysep\NAT@spacechar\NAT@hyper@{\NAT@date}}{}{}
  \patchcmd{\NAT@citex}
    {\@citea\NAT@hyper@{%
       \NAT@nmfmt{\NAT@nm}%
       \hyper@natlinkbreak{\NAT@spacechar\NAT@@open\if*#1*\else#1\NAT@spacechar\fi}%
         {\@citeb\@extra@b@citeb}%
       \NAT@date}}
    {\@citea\NAT@nmfmt{\NAT@nm}%
     \NAT@spacechar\NAT@@open\if*#1*\else#1\NAT@spacechar\fi\NAT@hyper@{\NAT@date}}
    {}{}
  \makeatother
\fi

\MakeObj[TVLM~513]{tvlm}{TVLM\%20513-46546}{TVLM~513--46546}
\MakeObj{lp349}{LP\%20349-25}{LP~349--25~AB}
\MakeObj{3c286}{3C\%20286}{3C~286}
\MakeObj{4c23.41}{4C\%2023.41}{4C~23.41}
\MakeObj{sigmagem}{V*\%20sig\%20Gem}{$\sigma$~Gem}
\MakeObj{eqpeg}{V*\%20EQ\%20Peg}{EQ~Peg}
\MakeObj{d1048}{DENIS-P\%20J104814.9-395604}{DENIS-P~J104814.9-395604}
\MakeObj{n33370}{NLTT\%2033370}{NLTT~33370~AB}

\newcommand\amm{{\ensuremath{\alpha_\text{mm}}}}
\newcommand\apx{\ensuremath{\sim}}

\newcommand\citeeg[1]{\citep[\eg,][]{#1}}
\newcommand\dd{\ensuremath{\text{d}}}
\newcommand\eg{\textit{e.g.}}

\newcommand\ewha{{\ensuremath{\text{EW}_\ha}}}
\newcommand\ha{{\ensuremath{\text{H}\alpha}}}

\newcommand\kms{km~s$^{-1}$}
\newcommand\mj{M$_\text{J}$}
\newcommand\ms{M$_\odot$}
\newcommand\prot{\ensuremath{P_\text{rot}}}
\newcommand\rj{\ensuremath{\text{R}_\text{J}}}
\newcommand\tb{\ensuremath{T_\text{B}}}

\newcommand\ujy{$\mu$Jy}

\newcommand\vsi{\ensuremath{v \sin i}}

\begin{document}

\title{The first millimeter detection of a non-accreting ultracool dwarf}
\author{
  P.~K.~G. Williams\altaffilmark{1},
  S.~L. Casewell\altaffilmark{2},
  C.~R. Stark\altaffilmark{3},
  S.~P. Littlefair\altaffilmark{4},
  Ch. Helling\altaffilmark{5},
  E. Berger\altaffilmark{1}
}
\email{pwilliams@cfa.harvard.edu}
\altaffiltext{1}{Harvard-Smithsonian Center for Astrophysics, 60 Garden Street,
  Cambridge, MA 02138, USA}
\altaffiltext{2}{Department of Physics and Astronomy, University of Leicester,
  University Road, Leicester LE1 7RH, UK}
\altaffiltext{3}{Division of Computing and Mathematics, Abertay University,
  Dundee DD1 1HG, UK}
\altaffiltext{4}{Department of Physics and Astronomy, University of Sheffield,
  Sheffield S3 7RH, UK}
\altaffiltext{5}{SUPA, School of Physics and Astronomy, University of St
  Andrews, St Andrews KY16 9SS, UK}

\slugcomment{ApJ in press (accepted 2015 Nov 9)}
\shorttitle{First millimeter ultracool dwarf detection}
\shortauthors{Williams et al.{}}

\begin{abstract}
  The well-studied M9 dwarf TVLM~513--46546 is a rapid rotator ($\prot \apx
  2$~hr) hosting a stable, dipolar magnetic field of \apx3~kG surface
  strength. Here we report its detection with ALMA at 95~GHz at a mean flux
  density of $56 \pm 12$~\ujy, making it the first ultracool dwarf detected in
  the millimeter band, excluding young, disk-bearing objects. We also report
  flux density measurements from unpublished archival VLA data and new optical
  monitoring data from the Liverpool Telescope. The ALMA data are consistent
  with a power-law radio spectrum that extends continuously between centimeter
  and millimeter wavelengths. We argue that the emission is due to the
  synchrotron process, excluding thermal, free-free, and electron cyclotron
  maser emission as possible sources. During the interval of the ALMA
  observation that phases with the maximum of the object's optical
  variability, the flux density is higher at a \apx1.8$\sigma$ significance
  level. These early results show how ALMA opens a new window for studying the
  magnetic activity of ultracool dwarfs, particularly shedding light on the
  particle acceleration mechanism operating in their immediate surroundings.
\end{abstract}


\section{Introduction}
\label{s.intro}

It has long been recognized that both stars and giant planets possess magnetic
fields \citep{h1908, z58}. The Sun and the planets have very different
magnetic phenomenologies, however, making it unclear what kind of magnetic
activity --- if any \citep{mbs+02} --- should be expected for objects with
masses between these extremes. Only relatively recently have observations made
it clear that intermediate-mass objects can host vigorous magnetic activity
that has characteristics reminiscent of \textit{both} of these regimes
\citep{bbb+01, b06b, bbf+10, had+06, mbr12, wcb14, hlc+15}. The study of
magnetic activity in ultracool dwarfs \citep[stars and brown dwarfs with
  spectral types $\ge$M7;][]{krl+99, mdb+99} not only provides insight into
the internal structures and local environments of these objects
\citeeg{sksd12, nbc+12}, but also leads the way toward analogous studies of
exoplanets themselves, which have thus far eluded detection despite
significant efforts \citeeg{lf07, ldesgkz11}.

Radio observations have been the chief observational tool for understanding
magnetism in the ultracool regime. This is in part because other standard
tracers (\eg, X-ray and \ha\ emission) become difficult to measure in these
faint, rapidly-rotating objects \citep{smf+06, bbf+10, wcb14, gmr+00, shw+15},
but also because radio observations reveal an enticing and puzzling
phenomenology. Bright, highly-polarized, periodic pulses are interpreted as
auroral phenomena, specifically bursts due to the electron cyclotron maser
instability \citep[ECMI;][]{the.ecmi, t06} that diagnose large-scale
magneto/ionospheric current systems \citep{had+06, lmg15}. For reasons that
are poorly understood, however, pulse structures vary on both short and long
timescales, sometimes disappearing completely \citep{bgg+08, wb15}. Meanwhile,
the broadband non-flaring (``quasi-quiescent'') emission from these objects is
reminiscent of gyrosynchrotron emission from active stars, but often has an
unusually flat spectrum ($0 < \alpha < -0.5$, $S_\nu \propto \nu^\alpha$), can
have significant polarization, and seemingly occurs at a spectral luminosity
that is independent of bolometric luminosity \citep{opbh+09, mbi+11, had+08}.
In some sources this emission appears steady over \apx year timescales
\citep{opbh+09} while in others it varies significantly \citep[factor-of-two
  changes;][]{adh+07, adh+10}; to date radio monitoring has not been
consistent enough to shed light on the relevant evolutionary time scales,
which could potentially reflect magnetic activity cycles \citep{sb99}. Another
major puzzle is the origin of the energetic electrons driving the ECMI bursts;
two proposed sources are atmospheric ionization processes \citep{hjwd11} and
magneto-ionospheric coupling currents \citep{nbc+12}.

Modern radio telescopes are capable of achieving \apx\ujy\ sensitivities at
high frequencies ($\gtrsim$20~GHz), raising the possibility of probing the
means by which particles are accelerated to MeV energies by objects with
effective temperatures of $\lesssim$2500~K. Here we present a detection of the
ultracool dwarf \objf{tvlm} (hereafter \obj{tvlm}) at 95~GHz with the Atacama
Large Millimeter/submillimeter Array (ALMA), the first such result at
millimeter wavelengths.\footnote{Here and throughout this work we disregard
  millimeter detections of disks around young ultracool dwarfs
  \citeeg{kab+15}.}

We proceed by summarizing the characteristics of \obj{tvlm}
(\autoref{s.target}), then describing our observations and data reduction
(\autoref{s.obs}). We next discuss some intriguing features in the data
(\autoref{s.disc}). Finally we present our conclusions (\autoref{s.conc}).

\section{The ultracool dwarf TVLM 513--46546}
\label{s.target}

\obj{tvlm} was first determined to be a nearby, faint star by \citet{t93b}
through trigonometric parallax measurements. \citet{khs95} originally adopted
a spectral type of M8.5, but both \citet{rck+08} and \citet{wmb+11} update the
assignment to M9. Astrometric VLBI monitoring has yielded a precise
trigonometric parallax implying a distance of $10.762 \pm 0.027$~pc
\citep{fbr13} and an absolute $K_s$-band \citep[2MASS;][]{the.2mass} magnitude
of $10.547\pm0.025$~mag. A lack of Li absorption lines implies an age
$\gtrsim$100~Myr and, in connection, a mass $\gtrsim$0.06~\ms\ \citep{mrm94,
  rkl+02}, while membership in the ``young/old disk'' kinematic category of
\citet{l92} is suggested by a low space velocity \citep{lah98}. The VLBI
astrometry excludes the presence of unseen companions with masses
$\gtrsim$4~\mj\ at orbital periods $\gtrsim$10~d or with masses
$\gtrsim$0.3~\mj\ at periods $\gtrsim$710~d \citep{fbr13}. NIR imaging
excludes companions with separations between 0.1 and 15~arcsec \citep{csfb03}.

While it has relatively weak \ha\ emission \citep[$\ewha \apx
  2$~\AA;][]{mrm94, b01}, \obj{tvlm} is one of the most-observed radio-active
ultracool dwarfs \citep{b02, had+06, ohbr06, hbl+07, bgg+08, fb09, dam+10,
  jol+11, fbr13, wr14, lmg15}. Its non-flaring flux density varies at the 30\%
level over \apx year timescales \citep{adh+10}. Both \citet{ohbr06} and
\citet{hbl+07} report quasi-simultaneous flux density measurements of
\obj{tvlm} at multiple frequencies, providing insight as to the shape of its
broadband spectrum, as explored below.

Additionally, \obj{tvlm} usually emits periodic ECMI flares that allow
measurement of its rotation period, \prot, and surface field strength
\citep{had+06}. \citet{dam+10} used these flares to measure $\prot =
1.96733 \pm 0.00002$~hr, consistent with spectroscopic measurements of $\vsi
\apx 60$~\kms\ \citep{b01, mb03}. \obj{tvlm} is also optically variable at the
\apx10~mmag level in the $I$~band at the same periodicity \citep{lhz+07}. This
variability is recovered in the SDSS $i'$ and $g'$ bands with an
180\degr\ phase shift between the two \citep{ldm+08}, which was originally
interpreted as signifying the presence of a dust cloud but may be associated
with \obj{tvlm}'s known auroral phenomena \citep{hlc+15}. Five years of $I$
and $i'$ band monitoring by \citet{hhb+13} lead them to measure $\prot =
1.95958 \pm 0.00005$~hr with stable phasing over that time period. They find a
peak-to-peak variability amplitude that varies with time between \apx0.6 and
1.2\%. \citet{wr14} combine the radio and optical data to obtain a consistent
value, $\prot = 1.959574 \pm 0.000002$~hr. Most recently, \citet{x.mpzop15}
report periodic variation in the linear polarization of \obj{tvlm}'s optical
emission at a \apx65\degr\ phase difference from the total intensity,
suggesting the presence of free electrons close to the object ($d \lesssim
R_*$) or inhomogeneous dust structures, as suggested by \citet{ldm+08}.

\section{Observations and Data Reduction}
\label{s.obs}

\subsection{ALMA}
\label{s.obs.alma}

\begin{figure}[tb]
  \plotone{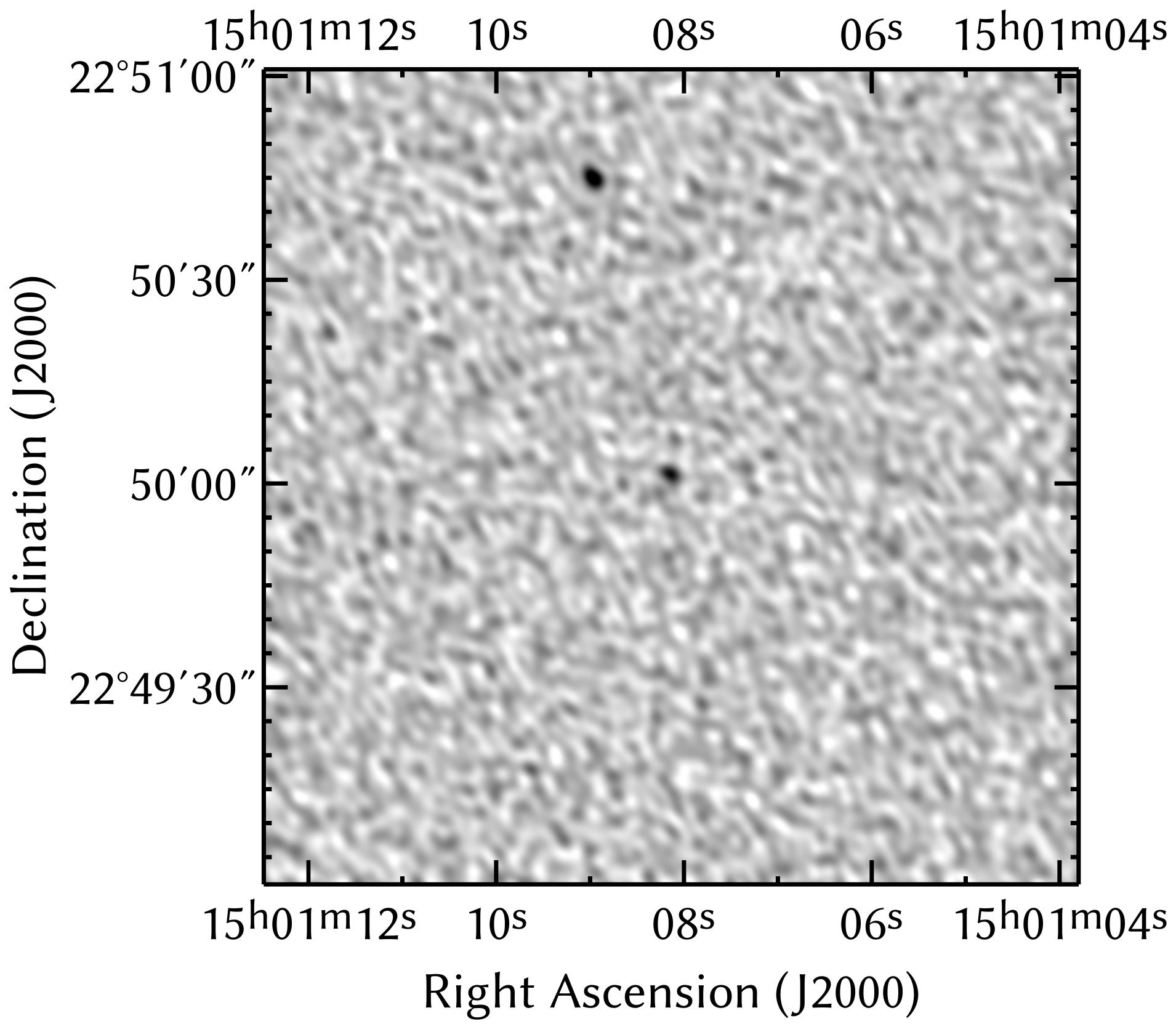}
  \caption{ALMA detection image of \obj{tvlm}. The effective frequency of the
    image is 97.48~GHz, the total bandwidth is 8~GHz, the grayscale runs from
    $-20$~\ujy\ (white) to 65~\ujy\ (black), and the image rms is
    \apx8.4~\ujy. A source is detected at the expected position of \obj{tvlm}
    with a flux density of $55 \pm 12$~\ujy, where the uncertainty in the flux
    density comes from least-squares fitting the pixel data. The synthesized
    beam is $3.1 \times 2.2$~arcsec at a position angle of 44\degr.}
  \label{f.image}
\end{figure}

\obj{tvlm} was observed with ALMA on UT date 2015~Apr~03 (proposal
2013.1.00293.S, PI: Casewell) using the ``band 3'' receivers (\apx100~GHz
observing frequency) in the standard wideband continuum mode. The correlator
recorded data for two sidebands centered on 91.46~GHz (lower sideband, LSB)
and 103.49~GHz (upper sideband, USB), each having a total bandwidth of 4 GHz.
The observations lasted just under 4 hours, with the total on-source
integration time being 2.15~hr. Full Stokes correlations were not obtained,
precluding a polarimetric analysis.

Calibrated visibilities and a preliminary image were provided to us by the
European ALMA Regional Centre (EARC) after reduction by their staff. We
examined the data and identified no lingering problems. \autoref{f.image}
shows the result of our re-imaging of the full data set, using a somewhat
larger field of view than the EARC analysis.

Using the astrometric parameters reported by \citet{fbr13}, the expected
position of \obj{tvlm} at the time of the ALMA observations was RA =
15:01:08.1463, Dec. = $+22$:50:01.224, with an approximate uncertainty of
1~mas. In the ALMA image we find a source that is 6.2$\sigma$ above the
background noise at a position of RA = 15:01:08.139, Dec. = $+22$:50:01.14.
Fitting a source model to the image, we measure a flux density of $56 \pm
12$~\ujy\ and a positional uncertainty of 0.5~arcsec, where the uncertainty on
the flux density is higher than the image rms (9~\ujy) because it includes
covariances with the source positions. The positions are consistent to
0.2$\sigma$, and we identify this source with \obj{tvlm}.

\begin{figure}[tb]
  \plotone{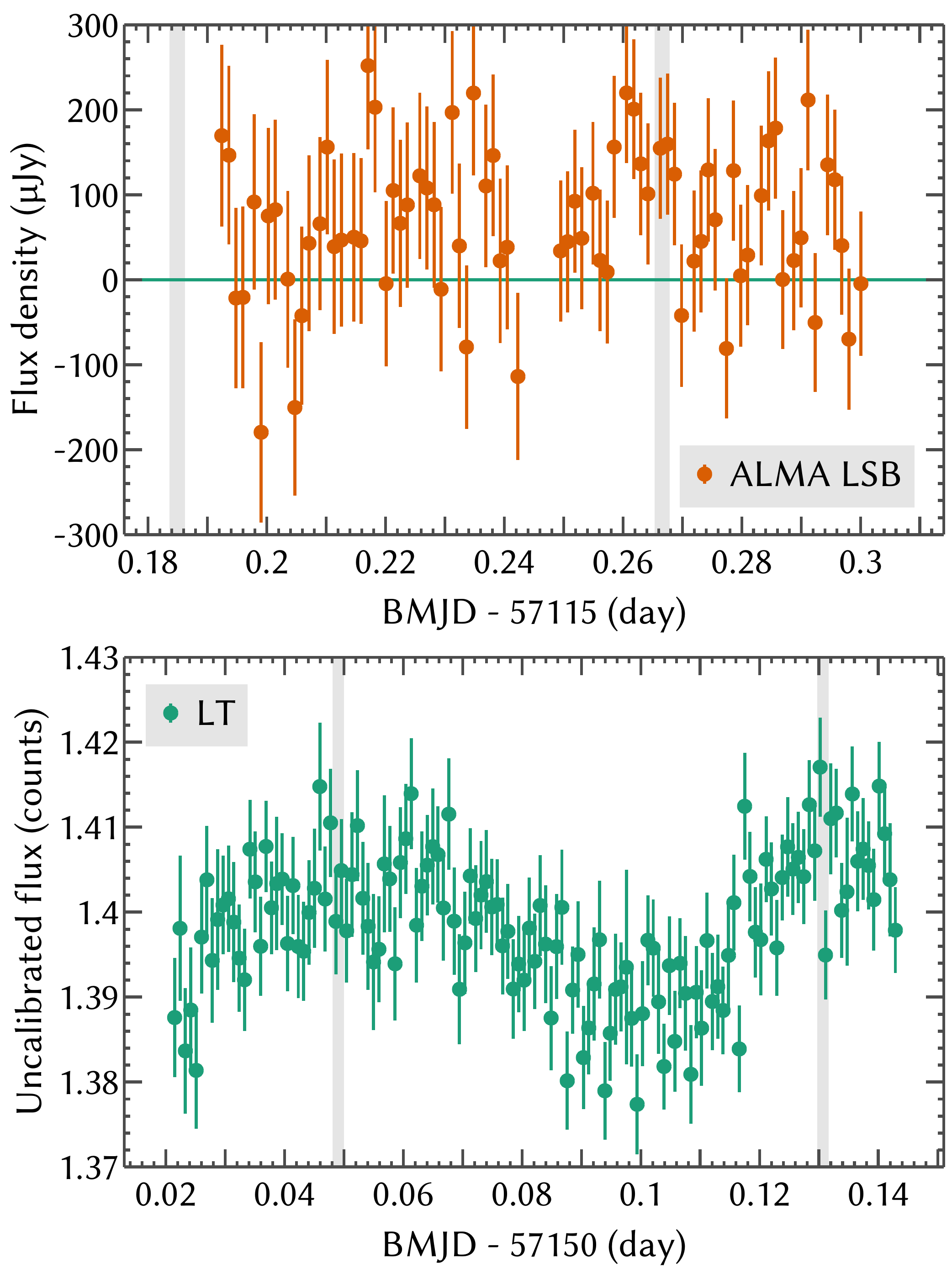}
  \caption{Light curves of \obj{tvlm} in millimeter (ALMA LSB; upper panel)
    and optical (lower panel) bands. The sampling intervals are 102 and 78
    seconds, respectively. Both panels have widths of 3.31~hr. The vertical
    gray bands indicate the 1$\sigma$ confidence bounds on the time of optical
    maximum as derived empirically from the LT data and propagated assuming a
    period of $1.95958 \pm 0.00005$~hr \citep{hhb+13}. The ALMA observations
    occurred 426 rotations prior to the LT observations. The possible increase
    in millimeter emission at BMJD 57115.265 occurs at the time of optical
    maximum.}
  \label{f.lightcurves}
\end{figure}

We imaged the two ALMA sidebands separately to assess the spectral index of
\obj{tvlm} at millimeter wavelengths. Our results are summarized in
\autoref{t.fluxdensities}. In the LSB image, with an effective frequency of
91.46~GHz, the flux density of the source is $66 \pm 8$~\ujy. In the USB
image, with an effective frequency of 103.49~GHz, it is $31\pm18$~\ujy. The
rms of the latter image is \apx12~\ujy, so that the signal is at the
2.6$\sigma$ level. The spectral index inferred from the ALMA data alone is
$\amm \approx -5 \pm 4$, largely unconstrained but inconsistent at the 97\%
confidence level with thermal emission, defined here as $\alpha \ge 1.5$
\citep{aw05}.

We extracted light curves from the calibrated LSB, USB, and band-averaged ALMA
visibilities using the technique described in \citet{wbz13}, subtracting the
lone additional source visible in the image from the visibilities. The
resulting raw light curves have a sampling cadence of 2~s. The top panel of
\autoref{f.lightcurves} shows the LSB light curve after averaging to a bin
size of 102~s, splitting bins across individual scans. We are unable to find
conclusive evidence for variability, although there is a possible flux density
increase around BMJD~57115.265 that we discuss below.

\subsection{Archival VLA Data}
\label{s.obs.vla}

We also analyzed unpublished archival observations of \obj{tvlm} performed
with the Karl G. Jansky Very Large Array (VLA). \obj{tvlm} was observed at
three frequencies on UT date 2010~Dec~16 while the VLA upgrade was being
performed (project ID TRSR0033, PI: Hallinan). Three sequential observations
were performed in K~band (\apx20~GHz), C~band (\apx6~GHz), and L~band
(\apx1.5~GHz) over the course of 1~hr. In the K~band observation, data were
taken in two basebands with center frequencies of 19.0 and 20.5~GHz and
bandwidths of 1.024~GHz each. The C~band observation used the same
configuration with center frequencies of 4.8 and 7.4~GHz. The L~band
observation had two basebands of 0.512~MHz bandwidth configured contiguously,
resulting in an effective spectral window of width 1.024~GHz centered at
1.58~GHz. In all cases individual spectral channels were 2~kHz wide.
\obj{3c286} was the bandpass and flux density calibrator and \obj{4c23.41} was
the complex gain calibrator.

\begin{deluxetable}{rr@{}l@{\,}lcc}

\tablecolumns{6}
\tablewidth{0em}
\tablecaption{New Flux Density Measurements of \obj{tvlm}\label{t.fluxdensities}}
\tablehead{
\colhead{Frequency} & \multicolumn{3}{c}{Flux Density} & \colhead{Obs. Date} & \colhead{Facility} \\
\colhead{(GHz)} & \multicolumn{3}{c}{(\ujy)} & \colhead{(BMJD)} &  \\ \\
\multicolumn{1}{c}{(1)} & \multicolumn{3}{c}{(2)} & \multicolumn{1}{c}{(3)} & \multicolumn{1}{c}{(4)}
}
\startdata
1.57 &  & $217$ & $\pm 153$ & 55546.51 & VLA \\
6.10 &  & $214$ & $\pm 18$ & 55546.50 & VLA \\
19.75 &  & $136$ & $\pm 43$ & 55546.49 & VLA \\
91.46 &  & $66$ & $\pm 8$ & 57115.24 & ALMA \\
103.49 &  & $31$ & $\pm 18$ & 57115.24 & ALMA
\enddata
\end{deluxetable}

We calibrated the data using standard procedures in the CASA software system
\citep{the.casa}. We used the \textsf{aoflagger} tool to automatically flag
radio-frequency interference \citep{odbb+10, ovdgr12} and set the flux density
scale using the preliminary, 2010 version of the scale defined by
\citet{pb13}. From fitting a point-source model to images generated from data
in the three bands, we obtain the flux densities reported in
\autoref{t.fluxdensities}. The flux density measurement in L~band is marginal
but consistent with prior observations \citep{ohbr06}.

\subsection{Liverpool Telescope}
\label{s.obs.lt}

We observed \obj{tvlm} using IO:O on the fully robotic 2-m Liverpool Telescope
(LT) on La Palma, on the night of August 5th 2015 (UT). We used the Sloan
Digital Sky Survey (SDSS) $i'$ filter with exposure times of 60~seconds. We
bias-subtracted the images using the CCD overscan region and used twilight sky
flats for flat field correction.

We extracted fluxes for \obj{tvlm} and four nearby comparison stars using
apertures whose sizes were scaled to 1.6 times the measured seeing in each
image. Average seeing was 2\arcsec, and the seeing ranged between 1.6 and
4\arcsec. Conditions were photometric and all data were taken below an airmass
of 1.15. We used the brightest two comparison stars to correct the data for
transparency variations. We plot the LT optical light curve in the lower panel
of \autoref{f.lightcurves}.

We measured the time of optical maximum in the LT lightcurve to be
$\text{BMJD} = 57150.049 \pm 0.001$ by fitting a sinusoid to the data,
constraining the period to match the value determined by \citet{wr14}. The
combined radio/optical ephemeris described in that work predicts an optical
maximum at $\text{BMJD} = 57150.028$ (A.~Wolszczan, 2015, priv. comm.),
\apx90\degr\ out of phase with our observations.

\section{Discussion}
\label{s.disc}

Millimeter detections of stars are almost universally obtained from hot stars
with winds and/or shells, young stars with disks, or giants with extremely
large (\apx AU) photospheres at mm wavelengths \citep{atw94, pw96}. While
coronal-type emission has been suggested as a target for mm observations, we
are able to locate only a few robust detections at $\nu \gtrsim 90$~GHz in the
literature; these are of either active binaries and/or flares \citep{bpb+03,
  bb06, shb08}. We can locate no such published detections of single dMe flare
stars, although to the best of our knowledge the lack of ALMA detections of
these sources is due to a lack of observations rather than insufficient
sensitivity.

\subsection{Emission Mechanism}

The $\gtrsim$100~Myr age implied by the lack of Li absorption in \obj{tvlm}
\citep{mrm94, rkl+02} is significantly longer than typical disk dissipation
timescales of \apx10~Myr \citep[although there is evidence that disk lifetimes
  are inversely correlated with {[}sub{]}stellar mass; e.g.,][]{wc11}.
Furthermore, \obj{tvlm} does not have reported IR excesses, and the spectral
index of the ALMA data is not consistent with the $\alpha \approx 2$ expected
from thermal (photospheric or disk) emission \citep{rm97, aw05}. We therefore
reject this origin for the ALMA emission.

\citet{tm82} consider the relationship between X-ray emission and free-free
radio emission from the active M4+M5 binary \obj{eqpeg}. Assuming
parameters appropriate for active low-mass stars, they find that
$S_{\nu,\text{ff}} = (10^6\text{ Jy}) f_X$, where $f_X$ is measured over the
0.2--4~keV band in units of erg~s$^{-1}$~cm$^{-2}$. Applying \obj{tvlm}'s
measured X-ray flux of \apx$6 \times 10^{-16}$~erg~s$^{-1}$~cm$^{-2}$
\citep{bgg+08} and ignoring minor factors such as the chosen X-ray band, we
estimate a negligible free-free contribution of $S_{\nu,\text{ff}} \approx
1$~nJy. While this value is calculated for cm wavelength emission, the flat
spectrum associated with optically thin free-free emission implies that it
holds at mm wavelengths as well.

ECMI emission occurs predominantly at the cyclotron frequency $\nu_c = e B / 2
\pi m_e c$, requiring the presence of a magnetic field of $\gtrsim$34~kG
strength to explain the mm detections. This would be an order of magnitude
stronger than any directly-measured magnetic fields in low-mass stars
\citep{rb07, rb10, mdp+10} and exceeds theoretical limits on the surface field
strengths associated with fully convective dynamos of \apx3~kG \citep[from
  energy equipartition with convective motions;][]{fc14}. Therefore we reject
an ECMI origin for this emission.

Having eliminated these other options, we conclude that the observed ALMA
emission is synchrotron emission associated with \obj{tvlm}'s magnetic
activity, as detailed below. Our results represent one of a handful of
  detections of such emission from single dwarfs at ALMA frequencies
(84--950~GHz), ultracool or not \citep{bpb+03}. If additional observations
constrain the brightness temperature of the mm emission to be implausibly
large for the gyrosynchrotron process (e.g. detection of a rapid, bright
burst), the previous paragraph implies that non-ECMI coherent processes such
as the antenna mechanism \citeeg{k70} or plasma emission \citeeg{gz58} should
be considered.

\subsection{The Broadband Radio Spectrum of TVLM 513}
\label{s.disc.spectrum}

\begin{figure}[tb]
  \plotone{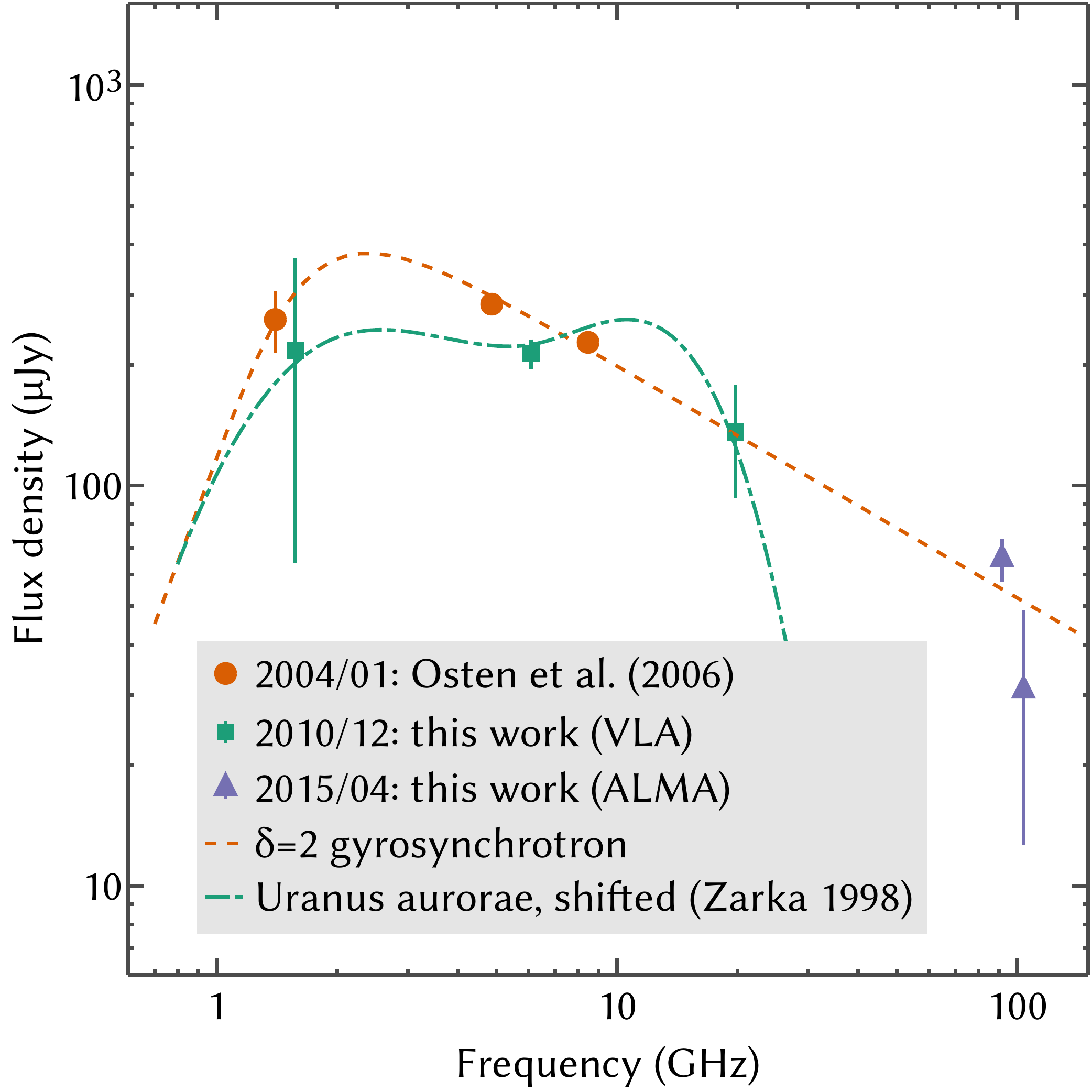}
  \caption{Non-simultaneous centimeter-to-millimeter spectrum of \obj{tvlm}.
    The orange points show the data obtained quasi-simultaneously by
    \citet{ohbr06}; green show the quasi-simultaneous VLA data described in
    \autoref{s.obs.vla}; blue show the simultaneous ALMA data described in
    \autoref{s.obs.alma}. The dashed lines show two representative spectral
    models described in \autoref{s.disc.spectrum}; they disregard the long
    time baseline between the relevant VLA and ALMA observations and should
    not be used for quantitative analysis.}
  \label{f.spectrum}
\end{figure}

In \autoref{f.spectrum} we assemble a composite centimeter-to-millimeter
spectrum of \obj{tvlm} from available simultaneous and quasi-simultaneous
data. All published radio observations of \obj{tvlm} lie within this frequency
range with the exception of a 2.5$\sigma$ upper limit of $S_\nu <
795$~\ujy\ at 325~MHz \citep{jol+11}. \obj{tvlm} is variable at the \apx30\%
level over long ($\gtrsim$month) timescales \citep{adh+10}, so that inferences
about the radio spectrum made by combining non-contemporaneous observations
should be treated cautiously. The results of \citet{ohbr06} and the VLA
analysis we present both suggest a spectrum peaking between 2 and 4 GHz with
$\alpha \approx -0.4$ at higher frequencies. In constrast, \citet{hbl+07}
report quasi-simultaneous observations at 4.88 and 8.44~GHz indicating a
rising spectrum. However, their observations were made one day apart, so that
variability may play a role in this finding. Furthermore, the flux densities
they report average over several bright, polarized flares that are
significantly brighter at the higher frequency. Using their Figure~3 to
estimate the bias caused by these flares, we infer that the non-flaring
spectrum in their observations has $\alpha \approx -0.1$.

In \autoref{f.spectrum} we show two possible models for the broadband spectrum
of \obj{tvlm}. We emphasize that these examples connect observations made
years apart and so should not be used for quantitative analysis. In orange, we
show a standard gyrosynchrotron model based on the equations of \citet{d85},
fit to our ALMA data and the measurements of \citet{ohbr06}. Here we fixed $B
= 100$~G and the electron energy spectral index $\delta = 2$, where $\dd N /
\dd E \propto E^{-\delta}$, which leads to $\alpha = -0.5$ in the
optically-thin limit. Our ALMA data are consistent with a model in which the
optically-thin portion of spectrum extends all the way to the ALMA band. In
this picture there is a hint of a spectral cutoff in the ALMA band. From the
ALMA measurements, however, we calculate a \apx12\% chance that $\amm > -0.5$,
i.e. that the mm spectrum is at least as flat as that observed at cm
wavelengths. Fixing the known distance of \obj{tvlm} and the properties of our
LSB detection, we infer a brightness temperature $\tb \apx (10^{6.4}\text{ K})
(r / \rj)^{-2}$, where $r$ is the characteristic length of the emitting region
\citep{d85}. This is consistent with gyrosynchrotron emission in the optically
thin regime unless $r \lesssim 0.003$~\rj.

\citet{had+06, had+08} have suggested that the broadband, non-flaring radio
emission of ultracool dwarfs may be due to the ECMI rather than the
gyrosynchrotron process, if there is continuous particle acceleration across a
range of magnetic field strengths and propagation effects depolarize the
emission. \citet{opbh+09} have used the plasma properties inferred from
multi-band observations to argue against this interpretation for the binary
\obj{lp349}, and \citet{rhhc11} analyze the broadband (5--24~GHz) radio
  spectrum of \obj{d1048} to argue similarly. As argued above, our detection
challenges this interpretation for \obj{tvlm} as well. To represent the sort
of spectrum that might result from such an emission mechanism, in
\autoref{f.spectrum} we show a version of the spectrum of Uranus' kilometric
radiation (UKR) as shown by \citet{z98}, where both the emission frequencies
and luminosities have been scaled arbitrarily to overlap the VLA measurements
reported in \autoref{t.fluxdensities}. While this is just one possible
example, it is representative of Solar System observations \citep{z98}.
Although the preexisting data could have been consistent with such a spectral
shape, the ALMA data exclude it.

\subsection{Phased Millimeter and Optical Variability?}

We do not find statistically significant evidence for variability in the mm
emission. While the data show a \apx15-minute period of elevated emission
around $\text{BMJD} = 57115.265$, we cannot exclude the possibility that this
elevated period is a noise fluctuation, given the duration of the observation
and measurement scatter. A simple likelihood ratio test comparing a constant
model and one with an exponentially fading flare yields an empirical $p$-value
of \apx7\% (\apx1.8$\sigma$); i.e., in about 7\% of simulated realizations of
the constant model, the likelihood ratio of the best-fit flaring and constant
models is least as high as that observed. Therefore the constant model cannot
be rejected with confidence. An alternative Bayes factor analysis yields
compatible results. The best-fit flare model achieves a peak flux density
\apx220~\ujy\ above the quiescent level.

Confidence in this possible signal could be increased if it were periodic and
observed multiple times, but the ALMA observation spanned only \apx1.3
rotations of \obj{tvlm} and the timing of the putative event prevents this
check. If the flare is real, it phases up
with the maxima of \obj{tvlm}'s optical variability as revealed by our LT
observations (\autoref{f.lightcurves}). Additional ALMA observations covering
multiple rotation periods, paired with contemporaneous optical monitoring,
could easily resolve whether the mm emission is periodically variable and how
it might phase relative to the optical emission. If the possible variation is
real and periodic, it would be out of phase with the polarized cm bursts,
which differ in phase from the optical maxima by $0.41 \pm 0.02$ of a period
\citep{wr14}.

\section{Conclusions}
\label{s.conc}

Our detection of \obj{tvlm} with ALMA makes it the first non-accreting
ultracool dwarf to be detected at millimeter wavelengths. This detection
bolsters the case for a gyrosynchrotron origin for the non-flaring radio
emission of ultracool dwarfs, although in this case, as in previous work
without such a large lever arm on the spectral index \citep{ohbr06}, a
relatively shallow electron energy index $\delta = 2$ is inferred. At the
high frequencies probed by ALMA, synchrotron cooling times are likely
comparable to observational durations: an ambient field strength $B \gtrsim
40$~G would lead to a cooling time shorter than the span of time on-source
in this observation \citep[2.6~hr;][]{p85}. For comparison, analyses of
radio spectra have concluded $B \approx 100$~G in \obj{n33370}
\citep{mbi+11} and $70 < B/\text{G} < 260$ in \obj{d1048} \citep{rhhc11}.
\citet{ohbr06} determined $B \lesssim 500$~G in \obj{tvlm}. Approximating
synchrotron-emitting electrons to emit at the frequency $\gamma^2 \nu_c$,
the observed ALMA emission originates in electrons with $\gamma \apx 180 (B
/ \text{G})^{-1/2}$. Monitoring observations can therefore likely probe the
fine-grained time dependence of the mechanism that injects energetic particles
into TVLM's surroundings, about which little is currently known.

However, confident inferences based on the broadband radio spectrum of
\obj{tvlm} are precluded because the ALMA observations were not obtained
contemporaneously with observations at longer wavelengths, and \obj{tvlm}'s
radio luminosity, and possibly its radio spectral shape, are variable.
Additional support from the Joint ALMA Observatory to allow simultaneous
observations with other observatories would be highly valuable. We note that
the observations in this work were performed in ALMA's ``Band 3,'' which is
less subject to weather restrictions than its higher frequencies. Similarly,
there are hints that there may be a periodic component in the mm emission of
\obj{tvlm}. In this particular case, the precisely-measured rotation period
makes it relatively easy to phase-connect observations obtained in
widely-spaced epochs, but given the possibly rapid evolution of emission in
the ALMA band and a desire to investigate objects that are not as
well-characterized, support from the Observatory for longer continguous
observations would likewise be of great importance to future studies.

Finally, we note that these results could only be obtained thanks to ALMA's
superior sensitivity compared to any previous mm-band instrument. This first
detection marks the opening of a new window for intensive investigations of
the magnetic dynamo and particle acceleration processes that are jointly
responsible for the cm/mm emission of ultracool dwarfs.

\acknowledgments

We thank Aleksander Wolszczan for providing the ephemeris prediction for our
optical observations. This paper makes use of the following ALMA data:
ADS/JAO.ALMA\#2013.1.00293.S. ALMA is a partnership of ESO (representing its
member states), NSF (USA) and NINS (Japan), together with NRC (Canada), NSC
and ASIAA (Taiwan), and KASI (Republic of Korea), in cooperation with the
Republic of Chile. The Joint ALMA Observatory is operated by ESO, AUI/NRAO and
NAOJ. The VLA is operated by the National Radio Astronomy Observatory, a
facility of the National Science Foundation operated under cooperative
agreement by Associated Universities, Inc. The Liverpool Telescope is operated
on the island of La Palma by Liverpool John Moores University in the Spanish
Observatorio del Roque de los Muchachos of the Instituto de Astrofisica de
Canarias with financial support from the UK Science and Technology Facilities
Council. We acknowledge support for this work from the National Science
Foundation through Grant AST-1008361. ChH acknowledges support from the
European Community under the FP7 by the ERC starting grant 257431. This
research has made use of the SIMBAD database, operated at CDS, Strasbourg,
France; NASA's Astrophysics Data System; and Astropy, a community-developed
core Python package for astronomy \citep{the.astropy}.

Facilities: \facility{ALMA}, \facility{Karl G. Jansky Very Large Array},
\facility{Liverpool Telescope}.

\bibliographystyle{yahapj/yahapj}
\bibliography{paper}{}

\end{document}